\documentclass{sig-alternate}

\newtheorem{theorem}{\bf Theorem}[section]
\newtheorem{property}[theorem]{\bf Property}

\begin{document}
\title{Maximizing profit using recommender systems}
\numberofauthors{3}
\author{
\alignauthor
Aparna Das
 \\\affaddr{Brown University}
\alignauthor
Claire Mathieu 
 \\\affaddr{Brown University}
\alignauthor
Daniel Ricketts
 \\\affaddr{Brown University}
}
\global\addauflag=1  
\date{}
\maketitle

\begin{abstract} 
Traditional recommendation systems make recommendations based solely on the customer's past purchases, product ratings and demographic data without considering the profitability the items being recommended. In this work we study the question of how a vendor can directly incorporate the profitability of items into its recommender so as to maximize its expected profit while still providing accurate recommendations. Our approach uses the output of any traditional recommender system and adjust them according to item profitabilities. Our approach is parameterized so the vendor can control how much the recommendation incorporating profits can deviate from the traditional recommendation. We study our approach under two settings and show that it achieves approximately 22\% more profit than traditional recommendations.
\end{abstract}

\section{Introduction}
Recommendation Systems are important tools for major companies like Amazon, Netflix and Pandora.  Ruse a customer's demographic data, past purchases and past product ratings to predict how the customer will rate new products \cite{at05,lsy03, bhk98,rv97}. Recommender systems have been shown to help customers become aware of new products of interest, increases sales and encourages customers to return to the business for future purchases \cite{cw06,skr99}. Designing recommendation systems that can accurately predict customer ratings has generated much research and interest both in the academic and business communities. The Netflix prize was a manifestation of this interest \cite{netflixprize}. However, the majority of the work on recommender systems has not explicitly considered how the profitability of products could incorporated into the recommendations. An article published in Knowledge@Wharton claims that the actual Netflix recommendation system modifies its ratings to encourage consumers to order the more obscure movies which are presumably cheaper for Netflix to supply than major blockbusters\cite{wharton}. While Netflix does not, publicly reveal that it uses such methods, it seems natural for a business to incorporate the profitability of products into its recommendations. Recommendations are, in some sense, extremely targeted advertising disguised as helpful suggestions, and the explicit goal of advertising is to increase profit for the company.

In this paper, we study the question of how a vendor should incorporate the profitability of items into its recommendations so as to maximize its expected profits. A naive approach is to give the most profitable items the highest recommendations. Then these items would presumably be bought more often and the business would make more money. However this tactic has some obvious flaws. While the customer may initially follow the vendor's recommendation, she may find that she does not like the items as much as the vendor predicted. After only a few such experiences, she would realize that the vendor's recommendations do not accurately reflect her tastes. In the best case for the vendor, the customer would ignore the vendor's recommendations and continue her natural purchasing behavior while in the worst case she would lose trust not only in the vendor's recommendations but also in the vendor as a whole and take her business elsewhere. Thus incorporating the profitability of items into recommendations must be done carefully so that customer's trust is not compromised. 

A reasonable assumption is that as long as the vendor consistently presents recommendations that are similar enough to the customer's own ratings, the customer will maintain a high level of trust in the accuracy of the vendor's recommendations. For this reason, we use an established \emph{ similarity measure }  as a measure of trust. We assume that the vendor has access to a vector $\vec{c}$ giving the consumer's true ratings for items. The vendor's objective is to present a recommendation vector $\vec{r}$ to the customer which is within a certain threshold of similarity to $\vec{c}$, and maximizes the vendor's expected profit (by incorporating the profitability of items into the recommendation). Section \ref{sec:model} gives details of the model. 

One question that might arise is how the vendor can determine the customer's true ratings (i.e $\vec{c}$ ) if the ratings are for products that the consumer has not yet rated herself. We believe there is a large amount of research and activity in developing highly accurate recommendation systems that solve this problem and assume the predictions from these systems for well established customers are good approximations to the customer's true ratings. 

The idea that the customer will maintain high trust as long as the vendor's recommendations are within a threshold of similarity of her own ratings has empirical support. Hill et al. showed that users asked to rate the same item at different times supply different ratings \cite{hill95}. If there is some natural variability in the ratings that users supply to the vendor, then slight differences between the predicted ratings and the actual customer ratings should be insignificant to the customer.

Chen et al. also considered using the profitability of items explicitly in recommender systems \cite{chch08}, but they do not explicitly require a level of accuracy in their system. 

The outline of our paper is: section \ref{sec:model} define our model and problem, section \ref{sec:sim} describes the similarity measure and gives justification for its use as a measure of customer trust and  section \ref{sec:profit} describes our approach for maximizing expected profit in different scenarios.

\section{Model}\label{sec:model}
Let $n$ be the number of items being sold by the vendor. Let $\vec{c}$ and $\vec{r}$ be the vectors of length $n$ where the $i$th components denoted $c_i$and $r_i$ gives a rating for item $i$. All items are rated using numbers between zero and some maximum rating $m$.  

We focus on the scenario where the vendor is interacting with an established customer who the vendor would like to continue to do business with in the long term. We assume the vendor uses a recommendation system which make  good predictions about how such a customer rates items and that vector $\vec{c}$ gives these predictions \footnote{The customer does not necessarily know all the entries of $\vec{c}$ herself as she has not purchased all items.}. The vendor presents the customer with recommendation vector $\vec{r}$ to help her decide which item to purchase. The customer has a certain level of {\em trust} in the accuracy of the vendor's recommendation which she has developed from experience with the vendor's past recommendations. When choosing an item to purchase, the customer considers $\vec{r}$ based on how much trust she has in the vendor's recommendations. 

The exact influence that a customer's level of trust in the vendor's recommendations has on her purchasing behavior is too complex to model precisely. For this reason, we make the following simplifying assumptions to allow for a model of that behavior: First we assume that the consumer's trust is closely tied to how similar the vendor's recommendations are to the ratings she would give items. Second we assume that as long the similarity threshold is consistently exceeded by the vendor, the consumer will use the recommendations when deciding what items to purchase. Specifically we assume there is a function $T(\vec{r})$ that assigns a scalar value to the notion of similarity between $\vec{r}$ and $\vec{c}$ where higher values indicate greater similarity. Second, we assume that if for every $\vec{r}$ that the vendor presents to the customer, $T(\vec{r})$ meets or exceeds some threshold value $\tau$, then the customer's trust in the recommendation system will remain at a constant, significantly high level. Finally, we assume that if the customer's trust remains at this level, her purchasing decision will be solely a function of $\vec{r}$. The customer's level of trust and her subsequent purchasing behavior is unknown if the vendor presents $\vec{r}$ such that $T(\vec{r})<\tau$.

The intuition behind these assumptions is that, as long as the vendor recommendation consistently predicts ratings that are similar enough to how the customer would rate the items, the customer will maintain a certain level of trust in the accuracy of the recommendation system. As a result, she will use the vendor's recommendation as information when deciding which items to purchase. However if $\vec{r}$ is too far from $\vec{c}$, the customer loses trust the vendor's recommendations and no longer considers them when deciding which item to buy. In that case, her purchasing behavior is unclear.

The vendor's main goal is to maximize profit. However, to maintain customer trust he is required to present vector $\vec{r}$ such that $T(\vec{r})\geq \tau$ for some constant $\tau$. Denote $\vec{\varphi}(\vec{r})$ be a vector valued function whose $i$th component gives the probability that the customer will purchase item $i$ item at a given time step. The customer, at any step, can purchase zero or more items, so the components of $\vec{\varphi}(\vec{r})$ need not sum to one. Let $\vec{p}$ be the profit vector whose $i$th component gives the profit received when when item $i$ is purchased. The vendor's expected profit is given by $E_p=\vec{p}\cdot\vec{\varphi}(\vec{r})$. Formally our problem is to maximize the vendor's expected profit while maintaining a level $\tau$ of trust with the customer: 
\begin{equation}\label{eq:opt} \textbf{Max } \vec{p}\cdot\vec{\varphi}(\vec{r}) 
\quad \textbf{ s.t. } T(\vec{r})\geq \tau
\end{equation}

\section{Similarity measures for trust}\label{sec:sim}
In this section we argue that the {\em Dice coefficient}, which measures similarity between two vectors, is an appropriate measure of consumer trust and briefly discuss why some of the other common singularity measures are lacking.

\noindent
{\bf Dice coefficient.} We adopt the {\em Dice  coefficient} given in equation \ref{eq:dice}, to measure trust $T(\vec{r})$. Note that the Dice coefficient is a popular measure which has previously been used to measure recommendation accuracy~\cite{chch08, ck04, skkr00}. 

\begin{equation}\label{eq:dice}
Dice(\vec{r}) = \frac{2 \sum_{i} c_i \cdot r_i}{\sum_i c_i^2 + \sum_i r_i^2} = \frac{\vec{c} \cdot \vec{r}}{|| \vec{c} ||^2+ ||\vec{r}||^2}
\end{equation}
Above $||\vec{x}|| = \sqrt{\sum_i x_i^2}$ denotes the length of vector $x$. Normally the Dice coefficient is denoted as a function of the two vectors whose similarity is being measured but here we use we denote it as only a function of $\vec{r})$ to emphasize the fact that $\vec{c}$ is constant known to the vendor.  Let $\theta$ denote the angle between $\vec{c}$ and $\vec{r}$. Another way of stating the Dice coefficient and Equation \ref{eq:dice} is, \begin{equation}\label{eq:diceangle}
Dice(\vec{r}) = \cos(\theta) \cdot \frac{2||\vec{c}|| ||\vec{r}||}{||\vec{c}||^2 + ||\vec{r}||^2}
\end{equation} 
We now list some properties of the Dice coefficient that makes it a reasonable function to measure trust. See the appendix for proofs.
\begin{property}\label{prop:dice1}
$Dice(\vec{r})$ is always between zero and one. $Dice(\vec{r}) = 1$, if and only if $c_i=r_i$ for every item $i$. $Dice(\vec{r}) = 0$, if and only if $r_i=0$ on all items $i$ such that $c_i > 0$.
\end{property}
Thus the $Dice(\vec{r})$ is one only when $\vec{r}$ is in complete agreement with $\vec{c}$, and it is zero only when $\vec{r}$ disagrees with $\vec{c}$ on all relevant items.
  
\noindent
{\bf Jaccard measure.} The Jaccard similarity measure, given below, behaves similar to the Dice coefficient and is used widely in information retrieval and data mining \cite{sg00, kr99, bb06}. 
$$Jac(\vec{r}) = (\vec{c}\cdot\vec{r})/(||\vec{c}||^2 + ||r||^2 - \vec{c}\cdot \vec{r})$$  
It is another appropriate measure of consumer trust and all results extend to the setting where $T(\vec{r}) = Jac(\vec{r})$. 

\smallskip

\noindent
{\bf Cosine measure.} The cosine similarity measure, given in equation \ref{eq:cos}, is equal to the cosine of $\theta$ \footnote{Recall that $\theta$ is  the angle  between $\vec{c}$ and $\vec{r}$.} It is always between zero and one as no item is rated less than zero. 
\begin{equation}\label{eq:cos}
Cos(\vec{r}) = \cos(\theta) = \frac{\sum_{i} c_i r_i}{\sqrt{ \sum_{i} r_i^2 \sum_{i} c_i^2}}
\end{equation}
The cosine measure is not influenced by difference in the lengths of $\vec{r}$ and $\vec{c}$ and this is the main reason it seems unsuitable for measuring trust as demonstrated in the following example.

\noindent
{\em Example.} Suppose the items are rated between $1$ and $5$. Consider a picky customer who rates all items low and a vendor that recommends all items high, i.e $c_i = 1$ and $r_i = 5$ for all $i$. The cosine measure will be to one because $\theta = 0$ indicating that the picky customer has high trust for the vendor's recommendation, which is surly not the case. 

\smallskip

\noindent
{\bf Mean squared error and distance measures.} The mean squared error (MSE) measures the average dissimilarity between $\vec{r}$ and $\vec{c}$. To measure similarity we could use 1-MSE, Equation \ref{eq:mse}, which is always between zero and one. Note that $m$ denote the maximum possible rating.
\begin{equation}\label{eq:mse}
MSE(\vec{r}) = \frac{\sum_i ((c_i/m -r_i/m))^2}{n}
\end{equation}
The 1-MSE measure gives equal credit for agreements on items the customer dislikes as on agreements on items she prefers and this makes it unsuitable for measuring the trust as demonstrated by the example below. The This example also applies to other distance based similarity measures such as Euclidean distance and Manhattan distance. 

\noindent
{\em Example} Suppose the items are rated between $0$ and $5$. Let $\vec{c} = [5,5,5, 1, 1, 1, \ldots, 1]$ which represents a customer who rates a few items very high but dislikes most items. Consider the following recommendation $\vec{r} = [1, 1, 1, \ldots, 1, 5,5,5]$ where the vendor gives the highest ratings to a few items that the customer dislikes and gives the low ratings to all other items including the items the customer prefers. MSE is $\Theta(1/n)$ so that 1-MSE approaches $1$ as the number of items $n$ gets large.

\section{Profit Maximization}\label{sec:profit}
Using the Dice similarity coefficient defined in Section \ref{sec:sim} as the trust function the vendor's optimization problem is,
\begin{equation}\textbf{Max } E_p(\vec{r}) = \vec{p} \cdot \vec{\varphi}(\vec{r}) 
\textbf{ s.t. } Dice(\vec{r})=\frac{2 \sum_{i} c_i \cdot r_i}{\sum_i c_i^2 + \sum_i r_i^2}\geq \tau
\label{eq:genoptproblem} 
\end{equation}
We outline a general approach for solving Equation \ref{eq:genoptproblem} and then apply our technique on two different objective functions obtained by alternative definitions of $\vec{\phi}(\vec{r})$. We analyze how much profit the vendor gains by presenting the customer with recommendation $\vec{r}$ rather than $\vec{c}$. 

\subsection{General Approach}\label{sec:general}
Concepts from vector calculus give us a general approach for solving the vendor's maximization problem stated in Equation \ref{eq:opt}. Adding $1/\tau$ to both sides of the Dice constraint above and simplifying reveals that it is equivalent to,  
\begin{equation}\label{eq:dicecircle}
\sum_i \left(r_i-\frac{c_i}{\tau}\right)^2 \leq \left(\frac{1}{\tau^2}-1\right)\sum_i c_i^2
\end{equation}
As $\vec{c}$ is a constant for our setting, the feasible $\vec{r}$ for our problem lie in a region enclosed by an n-sphere with radius $\sqrt{(\frac{1}{\tau^2}-1)\sum_i c_i^2}$ which we refer to as the {\em Dice sphere}.

The general approach to solving maximization problem of Equation \ref{eq:genoptproblem} involves two parts. The first is to determine if there are any local maxima that lie strictly inside the Dice sphere i.e. which satisfy $Dice(\vec{r},\vec{c})> \tau$. The second part is to find the vector $\vec{r}$ that maximizes expected profit over all vectors on the surface of the Dice sphere, i.e. which satisfy $Dice(\vec{r},\vec{c})= \tau$. The largest of the local maxima in the sphere and the maximum on the surface is the global maximum.

The gradient of the objective function is zero at each local maximum inside the Dice sphere. Thus all solutions to $\nabla E_p=0$ are candidate vectors. Let $\vec{r_1},\vec{r_2},...,\vec{r_k}$ be the list of all vectors that satisfy this property. While these vectors could be local minima or saddle points instead of local maxima, it is not necessary to distinguish between them. It is only necessary to find the greatest value of $E_p(\vec{r_i})$ for all $i$ and compare it to the maximum of all vectors on the surface of the sphere. Let $\vec{r_{in}}$ denote the maxima vector inside the Dice sphere.

The maximum vector on the surface of the Dice sphere can be found using the method of Lagrange multipliers. The maximum valued vector $\vec{r_s}$ will satisfy $\nabla Dice(\vec{r_s},\vec{c})=\lambda \nabla E_p(\vec{r_s})$ and $Dice(\vec{r_s},\vec{c})= \tau$. The maximum of $\vec{r_s}$ and $\vec{r_{in}}$ is the solution to the optimization problem \cite{ch08, stewart}.

\subsection{Simple probability function}\label{sec:simpledot}
Consider a scenario where the customer can purchase zero or more items at each time step where the probability that the customer purchases item $i$ is independent of the vendor's ratings for other items. Here is a simple way to define the probability function which satisfies this assumption,
\begin{equation}\label{eq:linvarphi}
\vec{\varphi}(\vec{r}) = \left(\frac{r_1}{m}, \frac{r_2}{m},\ldots, \frac{r_n}{m} \right)
\end{equation}
Recall that $m$ is the highest possible rating for an item. With this definition of $\vec{\varphi}$, the probability that a customer purchases an item is linearly proportional to the vendor's rating of that item and the vendor's expected profit is
\begin{equation}\label{eq:linratingsEp}
E_p(\vec{r}) = \vec{p} \cdot \vec{\varphi}(\vec{r}) = \frac{1}{m}\sum_{i} p_i \cdot r_i
\end{equation}
If prices are all greater than zero, $\nabla E_p=(\frac{p_1}{m},...,\frac{p_n}{m})\neq 0$ so there are no local maxima inside the sphere. Thus we proceed to finding the maximum be on the surface of the Dice sphere.  Using Equation \ref{eq:dicecircle} we have $\nabla Dice(\vec{r},\vec{c})= \left( 2(r_1-\frac{c_1}{\tau}),...,2(r_n-\frac{c_n}{\tau})\right)$, so the maxima on the Dice sphere surface must satisfy the following system of equations, 
\begin{eqnarray*}\label{eq:lagrange}
&\frac{p_i}{m} =  2\lambda(r_i- c_i/\tau) \quad \textrm{ for all } i \\
\textrm{ and }
&\sum_i (r_i-c_i/\tau)^2 = \left(\frac{1}{\tau^2}-1\right)\sum_i c_i^2 
\end{eqnarray*}
where the last equation requires $\vec{r}$ to lie within the Dice sphere.  Solving the first set of equations we obtain that 
\begin{equation}\label{eq:riinitial}
r_i=\frac{p_i}{2m\lambda}+\frac{c_i}{\tau}
\end{equation}
Substituting in the value of $r_i$ into Equation \ref{eq:dicecircle} we get $\lambda=\frac{1}{2m}\sqrt{\frac{\sum_i p_i^2}{(\frac{1}{\tau^2}-1)\sum_i c_i^2}}$. The final solution is obtained by plugging $\lambda$ into Equation \ref{eq:riinitial}. 
\begin{equation}\label{eq:rifinal}
 r_i=p_i\sqrt{\frac{(\frac{1}{\tau^2}-1)\sum_j c_j^2}{\sum_j p_j^2}}+\frac{c_i}{\tau}
\end{equation}

\noindent
{\bf Profit Gains.} By presenting the customer with recommendation $\vec{r}$ derived in Equation \ref{eq:rifinal} the vendor earns expected profit $E_p(\vec{r}) = \frac{1}{m} \left( \sum_i p_i^2 \sqrt{({1}/{\tau^2-1})\sum_j c_j^2/\sum_j p_j^2} + p_ic_i/\tau \right)$, which is simplified via the Cauchy-Schwarz inequality to \footnote{ The Cauchy-Schwarz inequality is $\sum_i p_i^2 \sum_i c_i^2 \ge (\sum_i p_ic_i)^2$},
\begin{equation}\label{eq:eprofitr}
E_p(\vec{r}) \ge \left(\sqrt{\frac{1}{\tau^2}-1} + \frac{1}{\tau}\right) \frac{\sum_i p_ic_i}{m}.
\end{equation}
The expected profit from $\vec{c}$ is $E_p(\vec{c}) = (\sum_i p_ic_i)/m$. The vendor's profit gain from presenting $\vec{r}$ rather than $\vec{c}$ is,
$$\frac{E_p(\vec{r}) - E_p(\vec{c})}{E_p(\vec{c})} = \sqrt{\frac{1}{\tau^2} - 1} + \frac{1}{\tau} -1 \ge 2(1/\tau -1).$$
For instance, if the vendor presents recommendation vectors that are within similarity threshold $\tau = .9$ to $\vec{c}$, then in expectation he earns at least $2(10/9 - 1) > 22\%$ more profit presenting $\vec{r}$ rather than $\vec{c}$.   

\subsection{Simple distribution}\label{sec:ratingsdist}
Now we consider the scenario where at each time step the customer purchases only one item but chooses which item to purchase based on how its rating compares to ratings of other items. A simple way to model this is to assume that item $i$ is purchased with probability $r_i/\sum_j r_j$. Thus the customer first scales each item by the sum of the vendor's ratings, and then chooses uniformly among all offered items.  The expected profit will be  
\begin{equation}\label{eq:ratingsEp}
E_p(\vec{r}) = \vec{p} \cdot \vec{\varphi}(\vec{r}) = \sum_{i} \frac{p_i \cdot r_i}{\sum_{j} r_j}
\end{equation}


The local maxima inside the Dice sphere occur where the gradient of the expected profit is zero. The gradient is
$$\nabla E_p = \left\langle \frac{p_1 \sum_j r_j - r_1p_1}{\left(\sum_j r_j \right)^2}, \ldots, \frac{p_n \sum_j r_j - r_n \cdot p_n}{\left(\sum_j r_j \right)^2} \right\rangle$$ 
For $\nabla E_p = 0$, it must be the case that $\sum_j r_j = r_i$ for all $i$. Thus if there are at least two items with different ratings then there are no local maxima inside the Dice sphere. 

We move on to find local maxima on the surface of the Dice sphere. If we try to use the method of Lagrange Multipliers as before we end up having to solve the following system of equations for variables $\lambda$ and $r_i$ for all $i$:
\begin{eqnarray}
\frac{p_i \sum_j r_j - r_i \cdot p_i}{ \left(\sum_j r_j \right)^2} = 2\lambda (r_i - \frac{c_i}{\tau}) \textrm{ for all } i 
\textrm{ and } Dice(\vec{r}) \ge \tau
\end{eqnarray}
Unfortunately we do not know how to find a solution for this as the first set of equations involves $r_i$ for all $i$. Recall that in section \ref{sec:simpledot} the corresponding equations for finding the maxima on the Dice surface were each a function of only one $r_i$. This lead us to seek an alternative approach. 
We will reduce solving the optimization problem under the definition of $E_p$ given in Equation \ref{eq:ratingsEp}  to solving a series of simpler optimization problems on which we can effortlessly apply the method of Lagrange Multiplier. To do so, first consider the decision version of our problem: {\em Does there exist a $\vec{r}$ such that $E_p(\vec{r}) \ge V$ and $Dice(\vec{r}) \ge \tau$?}

Under Equation \ref{eq:ratingsEp}, having $E_p(\vec{r}) \ge V$ is equivalent to having $\sum_i(p_i-V)r_i \ge 0$. Thus the decision version of our problem is equivalent to solving the following maximization problem and checking that its solution has  $r_i \ge 0$ for all $i$,
\begin{equation}\label{eq:decisionproblem} \textbf{Max }  E_p(\vec{r}) =  \sum_{i} (p_i-V)r_i  \textbf{ s.t. } Dice(\vec{r})\geq \tau
\end{equation}  
Equation \ref{eq:decisionproblem} can be solved using the general approach outlined in Section \ref{sec:general}. The gradient $\nabla E_p(\vec{r}) = 0$ iff $p_i-V = 0$ for all $i$. Thus as long some item is not priced $V$, there are no local maxima for Equation \ref{eq:decisionproblem} inside the Dice sphere \footnote{If all item are priced $V$, all $\vec{r}$ yield expected profit $V$ and we could pick any $\vec{r}$ which lies inside the Dice sphere.}.  

To find the maxima on the surface of the sphere, we solve the following system of equations,
\begin{equation}\label{eq:ratingssystem}
p_i -V = 2\lambda (r_i - c_i/\tau)  \quad \textrm{ for all } i \quad \textrm{ and } \quad  Dice(\vec{r}) \ge \tau
\end{equation}
Note that Equation \ref{eq:ratingssystem} differ from Equation \ref{eq:lagrange} only by constants so the solution derived in Section \ref{sec:simpledot} shifted by constants is a solution for Equation \ref{eq:ratingssystem}. We get that, 
$$r_i = \frac{p_i - V}{2\lambda} + \frac{c_i}{\tau}  \quad \textrm{where}\quad \lambda = \frac{1}{2} \sqrt{\frac{\sum_i (p_i - V)^2}{(1/\tau^2-1)\sum_i c_i^2}}$$ 
If for all $i$, $r_i \ge 0$, we return a ``yes" for the solution of the decision problem and otherwise we return  ``no''. 

To find an a solution for the original optimization problem which is arbitrarily close to optimal, we can do binary search along a bounded interval of possible values of $E_p$, checking the existences of solutions using decision version algorithm described above.  Let $V_{\max} = \max_i p_i$. The initial binary search interval can be set to $[0, V_{\max}]$ since $E_p(\vec{r}) \le V_{\max}$ as the customer purchases one item per time step.  Each binary search step reduces the search interval by half and doing more and more binary search steps brings us closer and closer to the optimal solution for the optimization problem. Let $\delta < 1$. A solution which is within distance $V_{\max}\delta$ of the optimal can be found by performing $\log(\frac{V_{\max}}{\delta})$ binary search steps. For example, let $\vec{r}_{\star}$ denote the optimal solution. With $\delta =1/V_{\max}$, we can obtain an approximate solution $\vec{r}_a$ such that $E_p(\vec{r}_a)  + 1 \ge E_p(\vec{r}_{\star})$ in $\log(\frac{V_{max}}{\delta}) = \log(V_{max}^2) = O(\log V_{\max})$ binary search steps.   

\smallskip
\noindent
{\bf Profit Gains.} Thus we are able to find a near optimal solution to the optimization problem with $E_p(\vec{r})$ as defined in Equation \ref{eq:ratingsEp} by solving a series of optimization problems of the kind solved in Section \ref{sec:simpledot}. The profit gains analysis from section \ref{sec:simpledot} extends to the last ``yes'' solution obtained for a decision problem. However as this ``yes'' solution is near optimal for the original optimization problem,  the profit gains will be close to that from section \ref{sec:simpledot}.

\section{Conclusion}
Traditional recommendation systems do not incorporate the profitability of items into its recommendations. In this work we propose one of the first methods that balances maximizing vendor's profits while providing accurate recommendations . Our approach can supplement any traditional recommendation system  and allows the vendor to control how much the profit based recommendation should deviate from the traditional recommendation. We study our approach under two settings and show that the vendor can raise approximately 22\% more profit by allowing a 10\% deviation. 
Our work is a starting point and we hope it will simulate new research on incorporating profits into recommendation systems. 
 
\bibliographystyle{acm}
\bibliography{refs}

\begin{thebibliography}{10}

\bibitem{at05}
{\sc Adomavicius, G., and Tuzhilin, A.}
\newblock Toward the next generation of recommender systems: a survey of the
  state-of-the-art and possible extensions.
\newblock In {\em IEEE Trans. Know. Data. Eng.\/} (2005), IEEE, pp.~734--749.

\bibitem{bb06}
{\sc Berry, M., and Browne, M.}
\newblock {\em Lecture Notes in Data Mining}.
\newblock World Scientific, 2006.

\bibitem{ch08}
{\sc Cain, G., and Herod, J.}
\newblock {\em Multivariable Calculus}.
\newblock www.math.gatech.edu/~cain/notes/calculus.html, 1997.

\bibitem{chch08}
{\sc Chen, L., Hsu, F., Chen, M., and Y., H.}
\newblock Developing recommender systems with the consideration of product
  profitability for sellers.
\newblock In {\em Information Sciences\/} (2008), Elsevier, pp.~1032--1048.

\bibitem{cw06}
{\sc Chen, P., and Wu, S.}
\newblock How does collaborative filtering technology impact sales? empirical
  evidence from amazon.com. working paper.
\newblock Available at SSRN: http://ssrn.com/abstract=1002698, 2007.

\bibitem{ck04}
{\sc Cho, Y., and Kim, J.}
\newblock Applications of web usuage and product taxonomy to collaborative
  recommendations in e-commerce.
\newblock In {\em Expert systems with Applications\/} (2004), Elsevier,
  pp.~233--246.

\bibitem{hill95}
{\sc Hill, W., Stead, L., Rosenstein, M., and Furnas, G.}
\newblock Recommending and evaluating choices in a virtual community of use.
\newblock In {\em Proceedings of ACM CHI'95 Conference on Human Factors in
  Computing Systems\/} (1995), ACM Press, pp.~194--201.

\bibitem{bhk98}
{\sc J.S., B., Heckerman, D., and C, K.}
\newblock Empirical analysis of predictive algorithms for collaborative
  filtering.
\newblock In {\em Proceedings of the 14th Conference of Uncertainty in AI\/}
  (1998), Morgan Kaufmann Publishers, pp.~42--52.

\bibitem{kr99}
{\sc Kaufman, L., and Rousseeuw, P.}
\newblock {\em Finding Groups in Data, An Introduction to Cluster Analysis}.
\newblock Wiley, New York, 1990.

\bibitem{lsy03}
{\sc Linden, G., B., S., and York, J.}
\newblock Amazon.com recommendations:item-to-item collaborative filtering.
\newblock In {\em IEEE Internet Computing\/} (2003), IEEE, pp.~76--80.

\bibitem{netflixprize}
{\sc {Netflix, Inc.}}
\newblock Netflix prize., Dec 2008.

\bibitem{rv97}
{\sc Resnick, P., and Varian, H.~R.}
\newblock Recommender systems.
\newblock In {\em Communication of ACM\/} (1997), ACM Press, pp.~56--58.

\bibitem{skkr00}
{\sc Sarwar, B., Karypis, G., Konstan, J., and Riedl, J.}
\newblock Analysis of recommendation algorithms for e-commerce.
\newblock In {\em Proceedings of the 3rd ACM Conference on Electronic
  Commerce\/} (2000), ACM Press, pp.~158--167.

\bibitem{skr99}
{\sc Schafer, J., Konstan, J., and Riedl, J.}
\newblock Recommender systems in e-commerce.
\newblock In {\em Proceedings of the 2nd ACM Conference on Electronic
  Commerce\/} (1999), ACM Press, pp.~158--166.

\bibitem{stewart}
{\sc Stewart, J.}
\newblock {\em Multivariable Calculus 5e (Fifth Edition)}.
\newblock Thompson-Brooks/Cole, Canada, 2003.

\bibitem{sg00}
{\sc Strehl, A., and Ghosh, J.}
\newblock Value-based customer grouping from large retail data-sets.
\newblock In {\em Proceedings of SPIE Conference on Data Mining and Knowledge
  Discovery\/} (2000), pp.~33--42.

\bibitem{wharton}
{\sc {Wharton}}.
\newblock Reinforcing the blockbuster nature of media: The impact of online
  recommender, Dec 2008.

\end{thebibliography}

\appendix

\section*{Properties of the Dice coefficient}

\smallskip

\begin{proof}(Proof of Property \ref{prop:dice1})
\item $Dice(\vec{r}) \ge 0$ because no item is rated less than zero. Thus the numerator of Equation \ref{eq:dice} is always positive. Now suppose that  $Dice(\vec{r}) > 1$. By Equation \ref{eq:dice}, this implies that $0 > \sum_i c_i^2 - 2r_ic_i + r_i^2  = \sum_i (r_i-c_i)^2$ which is a contradiction. Thus $Dice(\vec{r}) \le 1$.

\item Suppose that $r_i = c_i$ for every $i$. Then using Equation \ref{eq:dice} $Dice(\vec{r}) = 1$.  To prove the other direction suppose that $Dice(\vec{r}) = 1$.  The formulation of the Dice coefficient given in Equation \ref{eq:diceangle} can be used to show that $r_i = c_i$ for all $i$.   Note that $0 \le \cos(\theta) \le 1$ as no item is rated less than zero. The second term of Equation \ref{eq:diceangle} is non-negative by definition and the following proof by contradiction shows it is at most $1$:  suppose $2||\vec{c}|| ||\vec{r}||/(||\vec{c}||^2+ ||\vec{r}||^2) > 1$ then $2||\vec{c}|| ||\vec{r}|| > ||\vec{c}||^2 + ||\vec{r}||^2$ implying that $0>(||\vec{c}||-||\vec{r}||)^2$ which is false.  Thus if $Dice(\vec{r}) = 1$ both terms of Equation \ref{eq:diceangle} must be $1$. As $\cos(\theta)=1$ the angle between $\vec{c}$ and $\vec{r}$ is zero and the second term being $1$ implies that $(||\vec{c}|| - ||\vec{r}||) = 0$ i.e that $\vec{r}$ and $\vec{c}$ have equal length.  Together this implies that $r_i=c_i$ for all $i$.
 
\item  Finally note that when $Dice(\vec{r}) =0$, the numerator of Equation \ref{eq:dice} is zero which implies that $r_i$ must be zero for each $i$ such that $c_i > 0$.
\end{proof}

\end{document}